\documentclass[10pt,conference]{IEEEtran}

\usepackage{etoolbox}
\makeatletter
\def\ps@headings{%
\def\@oddhead{\mbox{}\scriptsize\rightmark \hfil \thepage}%
\def\@evenhead{\scriptsize\thepage \hfil \leftmark\mbox{}}%
\def\@oddfoot{}%
\def\@evenfoot{}}
\makeatother
\usepackage{amssymb}
\usepackage{amsmath}
\usepackage{amsfonts}
\usepackage{graphicx}
\usepackage{algorithm}
\usepackage{algorithmic}
\usepackage{epstopdf}
\usepackage{cite}
\usepackage{amsmath,bm}
\usepackage{subfigure}
\usepackage{graphicx}
\usepackage{color}
\usepackage{enumerate}
\usepackage{subeqnarray}
\usepackage{cases}

\newtheorem{theorem}{\textbf{Theorem}}

\newtheorem{corollary}[theorem]{\textbf{Corollary}}

\newtheorem{lemma}{\textbf{Lemma}}

\newtheorem{remark}{\textbf{Remark}}

\newtheorem{Def}{Definition}
\newtheorem{Prob}{Problem}
\begin{document}
\title{Communication, Computing and Caching for Mobile VR Delivery: Modeling and Trade-off}
\author{Yaping Sun, Zhiyong Chen, Meixia Tao and Hui Liu\\
Department of Electronic Engineering, Shanghai Jiao Tong University,  Shanghai,  P. R. China\\
Email: \{yapingsun, zhiyongchen, mxtao, huiliu\}@sjtu.edu.cn}
\thanks{This paper was partially supported by the National Natural Science Foundation of China (Grant No. 61671291, 61528103, 61521062, 61420106008 and 61571299, 61329101), and Shanghai Key Laboratory Funding (STCSM15DZ2270400).}
\maketitle
\begin{abstract}
Mobile virtual reality (VR) delivery is gaining increasing attention from both industry and academia due to its ability to provide an immersive experience. However, achieving mobile VR delivery requires ultra-high transmission rate, deemed as a first killer application for 5G wireless networks. In this paper, in order to alleviate the traffic burden over wireless networks, we develop an implementation framework for mobile VR delivery by utilizing caching and computing capabilities of mobile VR device.
We then jointly optimize the caching and computation offloading policy for minimizing the required average transmission rate under the latency and local average energy consumption constraints. In a symmetric scenario, we obtain the optimal joint policy and the closed-form expression of the minimum average transmission rate. Accordingly, we analyze the tradeoff among communication, computing and caching, and then reveal analytically the fact that the communication overhead can be traded by the computing and caching capabilities of mobile VR device, and also what conditions must be met for it to happen.
Finally, we discuss the optimization problem in a heterogeneous scenario, and propose an efficient suboptimal algorithm with low computation complexity, which is shown to achieve good performance in the numerical results.
\end{abstract}

\section{Introduction and Framework}\label{I}
Virtual reality (VR) over wireless networks is gaining increasing attention recently, due to their ability to generate a digital experience at the full fidelity of human perception \cite{burst}. Recently, a market report forecasts that the data consumption from wireless VR headsets (smartphone-based or standalone) will grow by over $650\%$ in the next 4 years (2017-2021)\cite{VRmarket}. One major mobile VR service is streaming $360^\circ$ videos, which are currently available on some major video platforms, such as Youtube, Facebook, etc. However, currently available mobile VR can only provide narrow field of view (FOV), e.g., about $90^\circ$ FOV for Daydream View. Achieving high-quality mobile VR delivery is extremely urgent nowadays for a real immersive experience.






However, achieving mobile VR delivery leads to ultra-high transmission rate requirement (on the order of G bit/s), deemed as a first killer application for $5G$ wireless network \cite{whitepaper}. This is mainly due to the following two facts. First, a $360^\circ$ VR video consumes much larger data (on the order of Gigabytes), since a $360^\circ$ video provides larger-than-normal view angle (e.g., $110^\circ \times 90^\circ$ versus $30^\circ \times 20^\circ$) and higher-than-normal resolution (e.g., $8K$ versus $4K$) \cite{E} for immersive and good user experience. Secondly, $360^\circ$ video delivery requires extremely low latency (less than $20$ms) so as to avoid dizziness and nausea \cite{whitepaper}.

In order to tackle such challenges, researchers in both academia and industry make great efforts. First of all, at any given time, considering that any user only watches a portion of the $360^{\circ}$ video, namely FOV \cite{3D}, 
 the corresponding FOV is chosen to be transmitted instead of the entire panoramic video, thereby saving bandwidth significantly. 
 Then, by knowing user's FOV, researches on multi-view and tile-based  video streaming are being investigated \cite{tile1}. Moreover, in order to further improve the quality of experience for tile-based video streaming, motion-prediction-based transmission is also being studied based on dataset collected from real users \cite{fixation}.

In addition to the aforementioned work \cite{3D,tile1,fixation} which mainly focus on the VR video-level design, researches on investigating the opportunities for mobile VR delivery that can be potentially obtained via efficiently utilizing resources at the mobile edge network (MEN), i.e., communications, computing and caching (3C), are also drawing increasing attention \cite{E,Simone,zhang,J,Yang}. Specifically, \cite{E} illustrates the potential gain obtained from utilizing the resources at the MEN via simulation results. \cite{Simone} provides an explicit VR framework, based on which the insights on how to deliver $360^\circ$ video in mobile network are illustrated. \cite{zhang} analyzes the optimal subcarrier allocation to maximize VR QoS for streaming VR videos over heterogeneous small-cell network. \cite{J} formulates an optimization framework for VR video delivery in a cooperative multi-cell network. \cite{Yang} develops a communication-constrained MEC framework to reduce communication resource consumption via exploiting the computation and caching resources at mobile devices.

\begin{figure}[t]
\begin{center}
 \includegraphics[width=6.5cm]{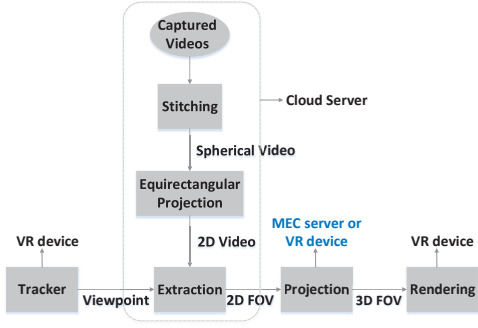}
\end{center}
 \caption{\small{A typical framework of VR service \cite{Simone}.}}
\label{Architecture}
\end{figure}

In this paper, we try to fully exploit the mobile edge resources to alleviate the communication burden over the wireless network for $360^\circ$ VR video delivery, especially the computing and caching capabilities at the mobile VR device\cite{3C}. 
To illustrate the problem at hand, we first analyze a typical VR framework \cite{Simone}, as shown in Fig. \ref{Architecture}: 
$1)$ \emph{Stitching}, which obtains a spherical video by stitching the videos captured by a multi-camera array; $2)$ \emph{Equirectangular projection}, which obtains 2D video by unfolding the spherical video; $3)$ \emph{Extraction}, which extracts the 2D video to obtain 2D FOV corresponding to the viewpoint obtained from the tracker at mobile VR device;
$4)$ \emph{Projection}, which projects the corresponding 2D FOV into 3D FOV; 
$5)$ \emph{Rendering}, which renders the obtained 3D FOV on the display of mobile VR device. Obviously, the tracker and rendering components must be computed at the mobile VR device. In addition, we assume that the stitching and equirectangular projection components are computed offline at the cloud server. Meanwhile, the extraction component is computed online at the cloud server. In this way, it can not only release mobile edge computing (MEC) server or mobile VR device from heavy computation process, but also alleviate the traffic burden on the wireless link, since such three components require the entire $360^\circ$ video as their inputs.

A key observation is that the projection component can be computed either at the MEC server or the mobile VR device due to its low computational complexity \cite{3D} and increasing computation capability at the mobile edge network \cite{E}. Specifically, offloading projection to the mobile VR device can save at least half of the wireless traffic load compared with computing projection at the MEC server. This is mainly due to the fact that in order to create a stereoscopic vision, the projection component has to be computed twice (one for each eye) to obtain two slightly differing images \cite{3D}, 
and thus the data size of its output (i.e., 3D FOV) is at least twice larger than that of its input (i.e., 2D FOV). However, offloading projection to the mobile VR device may incur longer latency compared with computing projection at the MEC server, since the computation capability of the MEC server is relatively stronger than that of the mobile VR device. Thus, the offloading policy requires careful design.
In addition, some FOVs can be cached at the mobile VR device proactively during idle traffic period based on the predictability of user FOV request process, so as to reduce the traffic burden over the wireless link during playback of VR video \cite{3C1,Hao,kongtao}.

Based on this proposed framework, we formulate the joint caching and offloading policy optimization problem to minimize the average transmission rate under both latency and local energy consumption constraints in this paper.
Specifically, we consider a cache-enabled MEC system with one single-antenna MEC server (e.g., base station (BS)) serving one mobile user via a wireless link. For a symmetric scenario, the optimal joint caching and computing offloading policy is obtained by analyzing the problem structure. Then, based on the closed form expression of the minimum average transmission rate, we analyze the three-way tradeoff among communication, computing and caching, which provides useful guidelines for future mobile VR delivery design. At last, we discuss the optimization problem for a heterogeneous scenario, and propose an efficient suboptimal algorithm with low computation complexity, which is shown to achieve good performance in numerical results. 
\section{System Model}
\begin{figure}[t]
\begin{center}
 \includegraphics[width=7.5cm]{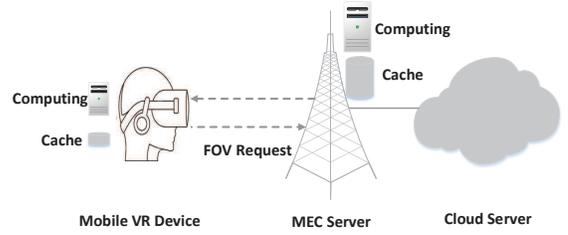}
\end{center}
 \caption{\small{System model.}
 }
\label{model}
\end{figure}

As illustrated in Fig.~\ref{model}, we consider a cache-enabled MEC system with one single-antenna MEC server (e.g., BS) serving one mobile user via a wireless link. The mobile user uses its single-antenna and cache-enabled mobile VR device to run a VR application. 
In this paper, we focus on the on-demand VR video streaming. As mentioned above, instead of transmitting the whole $360^{\circ}$ video, the MEC server only delivers the corresponding FOV to the mobile VR device at each time.


\subsection{VR Task Model}
For any viewpoint $i \in \mathcal{I}$, where $\mathcal{I} \triangleq \{1,..., N\}$ represents the viewpoint space \cite{J}, there exist a 2D FOV and a 3D FOV correspondingly, which are the input and output of the projection component as shown in Fig.~\ref{Architecture}, respectively. 
For each viewpoint $i\in \mathcal{I}$, the computation process of projecting the 2D FOV into 3D FOV is characterized
by a $4-$tuple $(I,O,w,\tau)$. Specifically, $I$ and $O$ denote the data size (\emph{in bits}) of the corresponding 2D FOV and 3D FOV, respectively. Generally, $O/I \geq 2$.  This is because in order to create a stereoscopic vision, the projection component has to be computed twice (one for each eye) to obtain two slightly differing images \cite{3D}.
$w$ and $\tau$ denote the required computation cycles per bit and the maximum tolerable latency (\emph{in seconds}), respectively. In addition, denote with $P_i$ the stationary probability for the mobile user requesting viewpoint $i$ \cite{J}. We consider uniform distribution, i.e., $P_i = \frac{1}{N}$ for each $i \in \mathcal{I}$.\footnote{The heterogeneous popularity distribution is considered in Section~\ref{heterogeneous}. In addition, in order to capture the temporal correlation of user's requests for different FOVs, we would like to model user request process for FOVs as a Markov chain in the future work.}



\subsection{Caching and Offloading}
First, we consider the cache placement at the mobile VR device. We assume that the mobile VR device is able to store the 2D FOVs for some viewpoints and equipped with a cache size $CI$ (\emph{in bits}), where $C$ is an integer.
Denote with $c_i \in \{0,1\}$ the caching decision for viewpoint $i$, where $c_i = 1$ means that the 2D FOV for viewpoint $i$ is cached in the VR device, and $c_i = 0$ otherwise.
Under the cache size constraint, we have
\begin{equation}\label{CacheSize}
\sum_{i=1}^N c_i \leq C.
\end{equation}

Next, we consider the computation offloading for the projection component. 
For each viewpoint $i\in \mathcal{I}$, denote with $d_i \in \{0,1\}$ the computation offloading decision, where $d_i = 0$ means that the projection component is computed at the MEC server and $d_i = 1$ means that the projection component is computed at the mobile VR device.

Specifically, if $d_i=0$, i.e., computing projection for viewpoint $i$ at the MEC server, the execution for delivering the requested 3D FOV consists of such two stages: 1) the 2D FOV downloaded from the cloud server is projected into 3D FOV at the MEC server; 2) the 3D FOV is transmitted to the mobile VR device via the wireless link. Denote with $R_i$ the transmission rate (\emph{in bit/s}) of the wireless link for viewpoint $i$ and $f_0$ the CPU-cycle frequency (\emph{in cycle/s}) of the MEC server. Then, the latency required for serving the requested 3D FOV is $\frac{Iw}{f_0} + \frac{O}{R_i}$. Under the latency constraint, i.e., $\Big(\frac{Iw}{f_0} + \frac{O}{R_i}\Big)(1-d_i) \leq \tau$, we have
\begin{equation}\label{latency0}
R_i \geq R^0(1-d_i),\ \ \ \ i \in \mathcal{I},
\end{equation}
where $R^0 \triangleq \frac{O}{\tau-\frac{Iw}{f_0}}$ represents the least required transmission rate when the projection is computed at the MEC server. Here, we assume that $\frac{Iw}{f_0} < \tau$ for feasibility, i.e., the projection computation at the MEC server can be completed within the deadline. Considering that the MEC server is usually connected to an electricity grid, we do not take into account the energy consumption at the MEC server in this paper.


If $d_i = 1$, i.e., offloading the projection component to the mobile VR device, the execution consists of such two stages: 1) the 2D FOV for viewpoint $i$ that is not cached in the mobile VR device is downloaded from the MEC server; 2) the projection component is computed at the mobile VR device. Denote with $f_1$ the CPU-cycle frequency (\emph{in cycle/s}) of the mobile VR device. Then, the required latency is $\frac{I(1-c_i)}{R_i} + \frac{Iw}{f_1}$. Under latency constraint, i.e., $\Big(\frac{I(1-c_i)}{R_i} + \frac{Iw}{f_1}\Big)d_i \leq \tau$, we have
\begin{equation}\label{latency1}
R_i \geq R^1(1-c_i)d_i,\ \ \ \ i \in \mathcal{I},
\end{equation}
where $R^1 \triangleq \frac{I}{\tau-\frac{Iw}{f_1}}$ represents the least required transmission rate when the projection is computed at the VR device and without caching. Here, we assume that $\frac{Iw}{f_1} < \tau$ for feasibility, i.e., the projection computation at the mobile VR device can be completed within the deadline.\footnote{It is reasonable considering that the computation capability of mobile VR device becomes stronger with the development of Graphic Processing Unit (GPU) as well as the projection component is not very complex.} Denote with $f_{min} \triangleq \frac{Iw}{\tau}$ the minimal required frequency for the mobile VR device. 
In brief, the latency constraints (\ref{latency0}) and (\ref{latency1}) can be rewritten as
\begin{equation}\label{bandwidth}
 R_i \geq R^0(1-d_i) + R^1(1-c_i)d_i,\ \ \ \ i \in \mathcal{I}.
\end{equation}
The energy consumption for computation per cycle at the mobile VR device with frequency $f_1$ is $k f_1^2$ , where $k$ is a constant related to the hardware architecture \cite{k}. 
Considering that the mobile VR device is powered by its own battery, denote with $\bar{E}$ the maximum average energy available at the VR device over a period of $\tau$. Then, under the average energy consumption constraint of the mobile VR device, we have
\begin{equation}\label{energy}
\frac{k f_1^2Iw}{N}\sum_{i=1}^N d_i \leq \bar{E},
\end{equation}
where $\frac{N\bar{E}}{kf_1^2Iw}$ is assumed to be an integer throughout this paper.
\section{Problem Formulation and Optimal Policy}
\subsection{Problem Formulation}
A feasible joint caching and offloading policy is defined as follows:
\begin{Def}[Feasible Joint Policy]
A feasible joint caching and offloading policy 
is a mapping from the system resource, i.e., $(C, \bar{E}, f_0, f_1)$, and the VR projection parameters, i.e., $(I,O,w,\tau)$, into the joint caching and offloading decision, denoted as $(\mathbf{c},\mathbf{d})$, where $\mathbf{c} \triangleq (c_i)_{i\in \mathcal{I}}$ denotes the caching decision, satisfying both the cache size constraint (\ref{CacheSize}) and the latency constraints (\ref{bandwidth}), and $\mathbf{d} \triangleq (d_i)_{i\in \mathcal{I}}$ denotes the offloading decision, satisfying both the latency constraints (\ref{bandwidth}) and the local average energy consumption constraint (\ref{energy}).
\end{Def}

Given a feasible joint policy, the average transmission rate is given as $\frac{1}{N}\sum_{i=1}^N  R_i$. In this paper, we would like to obtain the optimal joint policy to minimize the average transmission rate. 
 The optimization problem is formulated as follows:
\begin{Prob}[Joint Caching and Offloading Optimization]\label{Prob1}
\begin{align}
& \min_{{\mathbf{c},\mathbf{d}}} \ \ \ \ \frac{1}{N}\sum_{i=1}^N  R_i \nonumber\\
& \ \ s.t. \ \ \ \ (\ref{CacheSize})\ (\ref{bandwidth})\ (\ref{energy}). \nonumber
\end{align}
\end{Prob}

In order to minimize the average transmission rate, it is direct to observe that the latency constraint (\ref{bandwidth}) is reduced to equality, i.e., $R_i = R^0(1-d_i) + R^1(1-c_i)d_i$ for each $i \in \mathcal{I}$. 
 For ease of structural property analysis and without loss of equivalence, Problem \ref{Prob1} is rewritten as:
\begin{Prob}[Equivalent Joint Policy Optimization]\label{Prob2}
\begin{align}
& \min_{{\mathbf{c},\mathbf{d}}} \ \ \ \ R^0 - \frac{(R^0-R^1)}{N}\sum_{i=1}^N d_i - \frac{R^1}{N}\sum_{i=1}^N c_id_i  \nonumber\\
& \ \ s.t. \ \ \ \ \ \ \ \ \ \ \ \ \ \ \ \ \ \ \ \ \ (\ref{CacheSize})\ (\ref{energy}). \nonumber
\end{align}
\end{Prob}
Denote with $R^*$ the optimal objective value of Problem \ref{Prob2} and $(\mathbf{c}^*, \mathbf{d}^*$) the optimal joint decision.
\begin{remark}[Performance Gain Analysis]\label{gain}
From Problem \ref{Prob2}, we note that the first term of its objective function, i.e., $R^0$, represents the average transmission rate required without offloading, i.e., the projections for all the viewpoints are computed at the MEC server. The sum of the last two terms represents the local computing gain. Specifically, the second term of its objective function, i.e., $\frac{(R^0-R^1)}{N}\sum_{i=1}^N d_i$, represents the gain obtained from local computing without caching, which depends on the relationship between $R^0$ and $R^1$, and also the computing capacity of the mobile VR device. The last term, i.e., $\frac{R^1}{N}\sum_{i=1}^N c_id_i$, represents the gain obtained from local computing with caching, which depends on both the computation and caching capacity of the mobile VR device.
\end{remark}
\subsection{Optimal Joint Policy}
First, for notational simplicity, denote with $d \triangleq \sum_{i=1}^N d_i$ and $c \triangleq \sum_{i=1}^N c_i$ the number of offloaded projections and that of 2D FOVs cached at the mobile VR device, where from (\ref{energy}) and (\ref{CacheSize}), we have $d\in \{0,...,\frac{N\bar{E}}{k f_1^2Iw}\}$ and $c\in \{0,...,C\}$, respectively. Considering that the projection parameters of each viewpoint $i\in \mathcal{I}$ are the same, in the following discussion, without loss of generality, given $d$, we assume that the corresponding $\mathbf{d}=(d_i)_{i\in \mathcal{I}}$ is obtained from:
\begin{equation}\label{numOffload}
d_i =
\begin{cases}
1& \text{i = 1, ..., $d$,}\\
0& \text{otherwise.}
\end{cases}
\end{equation}

Next, by analyzing the objective function of Problem \ref{Prob2}, we obtain the following optimality property of the joint policy.
\begin{lemma}[Optimality Structural Property]\label{struc}
For any $i\in \mathcal{I}$ such that $d_i=0$, $c_i=0$ without loss of optimality.
\end{lemma}

\begin{remark}
Lemma~\ref{struc} indicates that caching 2D FOV of any viewpoint $i\in \mathcal{I}$ at the mobile VR device can achieve performance gain only when the corresponding viewpoint $i$ is computed at the mobile VR device. Otherwise, there is no need to cache it to save the storage cost of the VR device.
\end{remark}

Under the assumption of (\ref{numOffload}), from Lemma~\ref{struc}, we obtain the corresponding optimal caching decision $\mathbf{c}^*$ as follows.
\begin{corollary}[Optimal Caching]\label{optimalcaching}
Under the assumption\! (\ref{numOffload}), an optimal caching decision\! $\mathbf{c}^*$\! is given as
\begin{equation}\label{optcaching}
c_i^* =
\begin{cases}
1& \text{$i=1,...,C$,}\\
0& \text{otherwise,}
\end{cases}
\end{equation}
with $c^* = C$.
\end{corollary}
\begin{remark}[Optimal Caching] Based on (\ref{numOffload}) (\ref{optcaching}), the number of 2D FOVs cached at the mobile VR device can be independent of the number of offloaded tasks $d$ without loss of optimality. 
\end{remark}

Then, the objective function of Problem \ref{Prob2}, denoted as $f(d,\mathbf{c}^*)$, is rewritten as
\begin{equation}\label{objective1}
f(d,\mathbf{c}^*) \triangleq R^0 - \frac{R^0-R^1}{N}d - \frac{R^1}{N}\min\{d,C\}.
\end{equation}

Finally, by analyzing the structure of (\ref{objective1}), we obtain the optimal number of offloaded projections, denoted as $d^*$, and the corresponding minimum average transmission rate $R^*$ from the following lemma.

\begin{lemma}[Optimal Offloading and Minimum Average Transmission Rate]\label{optimaloffloading}
The optimal offloading decision $d^*$ is given as
\begin{equation}\label{optoffing}
d^* =
\begin{cases}
\frac{N\bar{E}}{k f_1^2Iw} ,& \text{if $\frac{N\bar{E}}{k f_1^2Iw} \leq C$},\\
\frac{N\bar{E}}{k f_1^2Iw} ,& \text{else if $R^0 > R^1$},\\
C, & \text{otherwise}.
\end{cases}
\end{equation}
And the corresponding optimal value $R^*$ is given as:
\begin{equation}\label{optB}
R^*=
\begin{cases}
R^0(1-\frac{\bar{E}}{k f_1^2Iw} ),& \text{if $\frac{N\bar{E}}{k f_1^2Iw} \leq C$}, \\
R^0-(R^0-R^1)\frac{\bar{E}}{k f_1^2Iw}-\frac{R^1C}{N},& \text{else if $R^0 > R^1$},\\
R^0(1-\frac{C}{N}), & \text{otherwise}.
\end{cases}
\end{equation}
\end{lemma}

\begin{remark}[How Many VR Tasks to be Offloaded?]
From (\ref{optoffing}), if $\frac{N\bar{E}}{k f_1^2Iw}\! \leq\! C$, $d^*$ is limited by the computation capacity at the VR device. This is because all the offloaded tasks can be cached at the VR device. Otherwise, if $R^0 > R^1$, $d^*$ is limited by the computation capacity at the mobile VR device, since offloading itself can save transmission rate. On the other hand, if $R^0 \leq R^1$, $d^*$ is limited by the local caching capacity $C$, since the gain of offloading only comes from local caching.
\end{remark}

\section{Discussions on Tradeoffs}
In this section, we discuss the tradeoff among communication, computing and caching, and the impacts of $f_1$ and $\bar{E}$ on the minimum average transmission rate $R^*$ based on Lemma~\ref{optimaloffloading}.

\subsection{Tradeoff among Communication, Computing and Caching}
\begin{figure}[t]
\begin{center}
 \includegraphics[width=5.5cm]{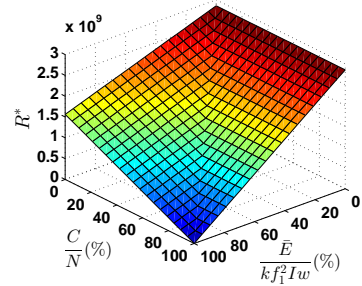}
\end{center}
 \caption{\small{ Tradeoff among communication, computing and caching under the latency constraint, where any ($R^*, \frac{C}{N}, \frac{\bar{E}}{kf_1^2Iw}$) point in this 3D figure can achieve $\tau=20$ ms. 
 $I = 25$M bits, $O = 50$M bits, $N = 6 \times 10^4$, $k = 10^{-27}$.}} 
\label{3C}
\end{figure}

Based on (\ref{optB}), we discuss the tradeoff among communication, computing and caching. Given $C$, we note that $R^*$ decreases with the computing capacity $\frac{N\bar{E}}{k f_1^2Iw}$. The performance gain obtained from the increase of $\frac{N\bar{E}}{k f_1^2Iw}$ increases with $C$, which grows from $46\%$ when $C =0$ to $ 100\%$ when $C=N$, as illustrated in Fig.~\ref{3C}.
This indicates that caching capacity promotes the exploitation of the computing capacity at the mobile VR device.
On the other hand, given $\frac{N\bar{E}}{k f_1^2Iw}$, we see that 
if $\frac{N\bar{E}}{k f_1^2Iw} > C$, $R^*$ decreases linearly with $C$. However, when $\frac{N\bar{E}}{k f_1^2Iw} \leq C$, $R^*$ does not decrease with $C$. Hence, given $\bar{E}$ and $f_1$, we obtain the minimum required cache size, denoted as $C^*$, which minimizes $R^*$, as follows.
\begin{corollary}[Minimum $C^*$ given $\bar{E}$ and $f_1$]
Given $\bar{E}$ and $f_1$, $C^* = \frac{N\bar{E}}{k f_1^2Iw}$ and $R^* = R^0(1-\frac{N\bar{E}}{k f_1^2Iw} )$, $\forall\  C \geq C^*$.
\end{corollary}

\begin{figure}[t]
\begin{center}
 \includegraphics[width=4.6cm]{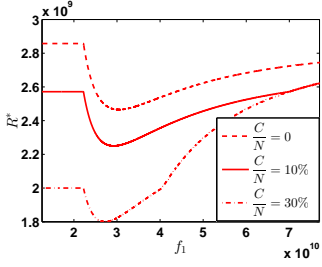}
\end{center}
 \caption{\small{ Impact of $f_1$ on $R^*$. Parameters are the same as that in Fig.~\ref{3C}.}}
\label{f_C}
\end{figure}
\begin{figure}[t]
\begin{center}
 \includegraphics[width=5.1cm]{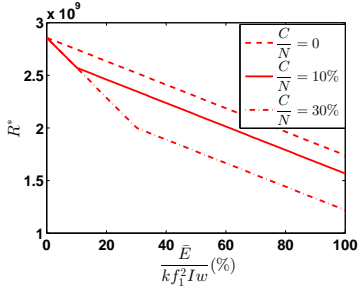}
\end{center}
 \caption{\small{ Impact of $\bar{E}$ on $R^*$. Parameters are the same as that in Fig.~\ref{3C}.}}
\label{E_C}
\end{figure}

\subsection{Impact of $f_1$ on $R^*$}

From (\ref{optB}), we observe that if $\frac{N\bar{E}}{k f_1^2Iw} \leq C$, $R^*$ increases with $f_1$ monotonically. This is mainly due to the fact that
 when $\frac{N\bar{E}}{k f_1^2Iw} \leq C$, the local computing gain with caching is limited by the computing capacity, while increasing $f_1$ decreases the computing capacity. Otherwise, if $R^0 \leq R^1$, $R^*$ does not decrease with $f_1$ but with $C$, since local computing without caching can not bring performance gain and the gain from offloading is limited by the caching capacity. On the other hand, if $R^0 > R^1$, by derivative w.r.t. $f_1$, we obtain the following lemma.
\begin{lemma}\label{intersection}
If $\frac{N\bar{E}}{k f_1^2Iw} > C$ and $R^0 > R^1$, $R^*$ first decreases and then increases with $f_1$.
\end{lemma}
\begin{remark}
Lemma~\ref{intersection} is mainly due to the fact that when $f_1$ is relatively small, increasing $f_1$ alleviates the transmission rate requirement, while when $f_1$ grows relatively large, increasing $f_1$ decreases the computation capacity.
\end{remark}


We illustrate the analytical result in Fig.~\ref{f_C}. We can observe that there exists an optimal computation frequency of the mobile VR device, denoted as $f_1^*$, which minimizes $R^*$. Extremely, when $C = 0$, via derivative w.r.t. $f_1$, we obtain $f_1^*$ as follows. 

\begin{corollary}[Optimal $f_1^*$ without Caching] When $C = 0$, $f_1^*$ that minimizes $R^*$ is obtained from
\vspace{-2mm}
\begin{equation}\label{f_1}
 f_1^* = \Big(1-\frac{I}{4R^0\tau} \Big)f_R + \sqrt{\Big(1-\frac{I}{4R^0\tau}\Big)^2f_R^2-\frac{Iw}{\tau}f_R},
\end{equation}
where $f_R \triangleq \frac{Iw}{\tau-I/{R^0}}$.
\end{corollary}

\subsection{Impact of $\bar{E}$ on $R^*$}
From (\ref{optB}), we observe that if $\frac{N\bar{E}}{k f_1^2Iw} \leq C$, $R^*$ decreases with $\bar{E}$ linearly at the decreasing rate $\frac{R^0}{kf_1^2Iw}$. Otherwise, if $R^0\geq R^1$, $R^*$ decreases with $\bar{E}$ linearly at the decreasing rate $\frac{R^0-R^1}{kf_1^2Iw}$, as illustrated in Fig.~\ref{E_C}. On the other hand, if $R^0 \leq R^1$, $R^*$ does not decrease with $\bar{E}$, since the gain of computing at the mobile VR device is only limited by the caching capacity.
\section{Heterogeneous Scenario Discussion}\label{heterogeneous}
In this section, we consider a heterogeneous scenario, where the projection parameters for each viewpoint may be different from those of others, denoted as $(I_i,O_i,w_i,\tau_i)$ for each viewpoint $i\in \mathcal{I}$. 
Similar to Problem~\ref{Prob2}, in order to obtain the optimal joint caching and offloading policy to minimize the average transmission rate, the optimization problem is formulated as follows:
\begin{Prob}[Joint Policy Optimization in Heterogeneous Scenario]\label{generalProb}
\begin{align}
 \min_{{\mathbf{c},\mathbf{d}}}\ \  & \sum_{i=1}^N P_iR_i^0 - \sum_{i=1}^N P_i(R_i^0-R_i^1)d_i - \sum_{i=1}^N P_iR_i^1c_id_i\nonumber \\
s.t.\ \  &\ \ \ \ \ \ \ \ \ \ \ \ \ \ \ \ \sum_{i=1}^N P_ikf_1^2I_iw_id_i \leq \bar{E}, \\
  \ \ \ & \ \ \ \ \ \ \ \ \ \ \ \ \ \ \ \ \sum_{i=1}^N I_ic_i \leq C' ,\\\nonumber
\end{align}
\end{Prob}
where $R_i^0 \triangleq \frac{O_i}{\tau_i-\frac{I_iw_i}{f_0}}$ and $R_i^1 \triangleq \frac{I_i}{\tau_i-\frac{I_iw_i}{f_1}}$ denote the minimum required transmission rate to satisfy the latency constraint, when the projection of viewpoint $i$ is computed at the MEC server and computed at the mobile VR device, respectively. $C'$ (\emph{in bits}) denotes the cache size at the VR device.

Note that Problem~\ref{generalProb} is equivalent to a bilinear knapsack problem. An optimal algorithm is proposed in \cite{bilinear}, which consists of two stages. The first stage is to implement a mountain climbing algorithm (MCA) to obtain a local maximum. The second one is to adjoin a cutting plane which eliminates a local maximum and yet does not eliminate any solution potentially better than the current incumbent. 
However, for the MCA in the first stage, at each iteration, optimizing the knapsack problem via dynamic programming yields an overall running time of $\mathcal{O}(N^2)$ and requires an overall memory space of $\mathcal{O}(N^2)$, respectively. Hence, it may be infeasible when $N$ is relatively large (e.g., $N= 10^4$). In order to tackle the computational complexity of MCA, heuristically, we propose a greedy algorithm (GA). In this way, the overall running time is reduced to $\mathcal{O}(NlogN)$ and the required memory space is of $\mathcal{O}(1)$ via heap sort. Note that GA is computed offline, i.e., before the mobile VR device runs the VR application. Thus, the quality of user experience will not be influenced by GA.

\begin{algorithm}[t]
\caption{Greedy Algorithm (GA)}
\label{GA}
\small{\begin{algorithmic}[1]
\STATE  \textbf{Joint Greedy Allocation}.
Sort $\mathcal{I}$ according to $\frac{P_iR_i^0}{I_i}$ in descending order and obtain $\mathbf{c}$ and $\mathbf{d}$ from
\begin{equation}
c_{\lfloor j \rfloor} =
\begin{cases}
1, & \text{$j = 1, \cdots, s_c-1$},\\
0, & \text{otherwise},
\end{cases}
\end{equation}
\begin{equation}
d_{\lfloor j \rfloor} =
\begin{cases}
1, & \text{$j = 1, \cdots, \min\{s_e-1,s_c-1\}$},\\
0, & \text{otherwise},
\end{cases}
\end{equation}
where $\lfloor j\rfloor$ represents the index $i\in \mathcal{I}$ with the $j$-th maximal value of $\frac{P_iR_i^0}{I_i}$, $s_c$ represents the split index satisfying $\sum_{j=1}^{s_c-1} I_{\lfloor j \rfloor} \leq C'$ and $\sum_{j=1}^{s_c} I_{\lfloor j \rfloor} > C'$ and $s_e$ represents the split index satisfying $\sum_{j=1}^{s_e-1} P_{\lfloor j \rfloor}kf_1^2I_{\lfloor j \rfloor}w_{\lfloor j \rfloor} \leq \bar{E}$ and $\sum_{j=1}^{s_e} P_{\lfloor j \rfloor}kf_1^2I_{\lfloor j \rfloor}w_{\lfloor j \rfloor} \!>\! \bar{E}$.

\STATE \textbf{Additional Offloading Greedy Allocation}. If $s_e>s_c$, sort $\mathcal{I}' \triangleq \{i\in \mathcal{I}:d_i = 0, R_i^0>R_i^1\}$ according to $\frac{R_i^0-R_i^1}{k I_iw_if_1^2}$ in descending order and obtain $\{d_i:i\in \mathcal{I}'\}$ from
\begin{equation}
d_{\lfloor j' \rfloor} =
\begin{cases}
1, & \text{$j' = 1, \cdots, s_e'$},\\
0, & \text{otherwise},
\end{cases}
\end{equation}
where $\lfloor j'\rfloor$ represents the index $i\in \mathcal{I}'$ with the $j'$-th maximal value of $\frac{R_i^0-R_i^1}{k I_iw_if_1^2}$ and $s_e'$ represents the split index satisfying $\sum_{j=1}^{s_e-1} P_{\lfloor j \rfloor}kf_1^2I_{\lfloor j \rfloor}w_{\lfloor j \rfloor}+\sum_{j'=1}^{s_e'-1} P_{\lfloor j' \rfloor}kf_1^2I_{\lfloor j' \rfloor}w_{\lfloor j' \rfloor}\leq \bar{E}$ and $\sum_{j=1}^{s_e-1} P_{\lfloor j \rfloor}kf_1^2I_{\lfloor j \rfloor}w_{\lfloor j \rfloor} + \sum_{j'=1}^{s_e'} P_{\lfloor j' \rfloor}kf_1^2I_{\lfloor j' \rfloor}w_{\lfloor j' \rfloor} > \bar{E}$.

\end{algorithmic}}
\end{algorithm}
\begin{figure}[t]
\begin{center}
 \includegraphics[width=5cm]{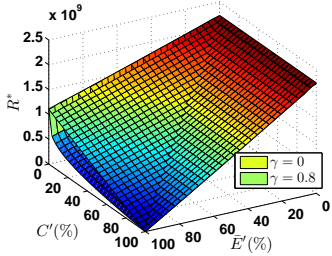}
\end{center}
 \caption{\small{ Tradeoff in Heterogeneous Scenario under GA. $\bar{C}' \triangleq \frac{C}{\sum_{i=1}^N I_i}$, $\bar{E}' \triangleq \frac{\bar{E}}{\sum_{i=1}^N P_ikf_1^2I_iw_i}$, $I_i \in [15,25]$ M bits and $P_i \propto \frac{1}{i^\gamma}$. The other parameters are the same as that in Fig.~\ref{3C}}.}
\label{3C1}
\end{figure}

As shown in Fig.\!~\ref{3C1}, in the heterogeneous scenario, the 3C tradeoff exhibits similar property to that in the symmetric scenario. Differently, when $\gamma\!\! =\! 0.8$, the gain obtained from the increase of \!$C$ \!decreases w.r.t.\! $C$, since as $\gamma$ grows, the probability that a requested file is requested and offloaded or cached becomes larger.

\section{Conclusion }

In this paper, we develop an implementation framework for mobile VR delivery by utilizing the caching and computing capacities of the mobile VR device and optimize the joint caching and offloading policy to minimize the average transmission rate under the latency and local energy consumption constraints in both symmetric and heterogeneous scenario. In the symmetric scenario,
we obtain the optimal joint policy and the closed form expression of the minimal transmission rate, based on which we analyze the tradeoff among communication, computing and caching. In the heterogeneous scenario, we propose GA, which is efficient and shown to achieve good performance.
In summary, we show theorectically that:
\begin{itemize}
  \item $R^*$ increases with $f_1$ if $\frac{N\bar{E}}{k f_1^2Iw} \leq C$;
  \item $R^*$ first decreases with $f_1$ and then increases with $f_1$ if $\frac{N\bar{E}}{k f_1^2Iw} > C$ and $R^0 \geq R^1$;
  \item $R^*$ stays unchanged with $f_1$ if $\frac{N\bar{E}}{k f_1^2Iw} > C$ and $R^0 < R^1$;
  \item $R^*$ decreases with the caching size $C$ of mobile VR linearly if $\frac{N\bar{E}}{k f_1^2Iw} > C$, while stays unchanged otherwise.
\end{itemize}
Note that in this paper, we mainly focus on obtaining first-order design insights and plan to further validate the proposed framework via real tests in future work.
\end{document}